# A new approach to the front-end readout of cryogenic ionization detectors: GeFRO

**C.Cattadori[a], B.Gallese[b], A.Giachero[a], C.Gotti[a,c,*], M.Maino[a], G.Pessina[a]**

[a] *INFN, Sezione di Milano-Bicocca and Dipartimento di Fisica dell'Università di Milano-Bicocca, P.za della Scienza 3, Milano 20126, Italy,*
[b] *Laboratori Nazionali del Gran Sasso, 67010, Assergi (L'Aquila), Italy,*
[c] *Dipartimento di Elettronica e TLC, Università di Firenze, Via S. Marta 3, 50125, Firenze Italy*

*E-mail*: Claudio.Gotti@mib.infn.it

ABSTRACT: We present GeFRO (Ge Front-end ReadOut) a novel approach to the readout of ionization detectors. The circuit allows to minimize the number of components and the space occupation close to the detector. This way a minimal impact is added on the radioactive background in those experiments where very low signal rates are expected, such as GERDA and MAJORANA. The circuit consists in a JFET transistor and a remote second stage. The DC feedback path is closed using a diode. Only two signal cables are necessary for biasing and readout.

KEYWORDS: Low Temperature Detectors, Low Background, Low Noise, Low Noise Amplifier.

[*] Corresponding author

## Contents

## 1. Introduction

The processing of signals from ionization detectors in very low background environment requires the minimization of the amount of mass contributed by the electronics equipment close to the detector area. Additional care is to be taken in minimizing power dissipation when the working environment is at cryogenic temperature, inside a refrigerator system. Examples of such set-ups are the experiments GERDA **[1]** and MAJORANA **[2]** both based on arrays of low background Ge detectors.

In the following sections we will describe in detail GeFRO (Ge Front-end ReadOut), our new approach aimed at minimizing the number of electronic components close to the detector.

## 2. The new front-end readout scheme for ionization detectors

The classical approach to the readout of an ionization detector is shown in Figure 1 **[3]**, **[4]**, **[5]**, **[6]**. The cascode/folded-cascode, Cas in the figure, which helps in optimizing the frequency bandwidth, is sometimes omitted in favor of a supply voltage reduction **[7**], **[8], [9], [10]**. The setup of Figure 1 works well provided that some precautions are adopted for the case the detectors are operated at cryogenic temperatures. Cables inside the refrigerator must have a small section for minimizing their thermal conductance, hence the heating injection. As a consequence, they are generally lossy and their series impedance could be responsible of non-negligible effects. Cross-talk can be injected from the positive supply $V_{CC}$ of Figure 1 if it is shared among many channels and the filtering capacitance $C_{F1}$ is missing. The terminated coaxial cable connected to the preamplifier output can be a few meters long and a loss in signal amplitude is verified if the preamplifier generates a voltage signal at its output.

When the mass close to the detector must be kept at minimum a few different configurations could be considered. The first is shown in Figure 2 **[11]**. The Charge Sensitive Preamplifier is split in two parts: the input stage, which consists of the JFET and the feedback capacitance and resistor, and the second stage located in a remote location and connected to the input stage with the transmission lines $T_A$ and $T_B$. Also this solution could be adequate, provided that attention is put on the effects of the phase shift, derived from the length of the connecting lines, that worsens the phase margin.

Another very simple and efficient approach is the voltage follower of Figure 3 **[12]**, **[13]**, **[14]**. Power consumption is minimized in this case since the biasing resistor is outside the refrigerator. Nevertheless an additional filtering capacitance, $C_{F1}$, is necessary to limit signal reflection at the drain of the JFET in fast signal applications, when long connecting lines are used.

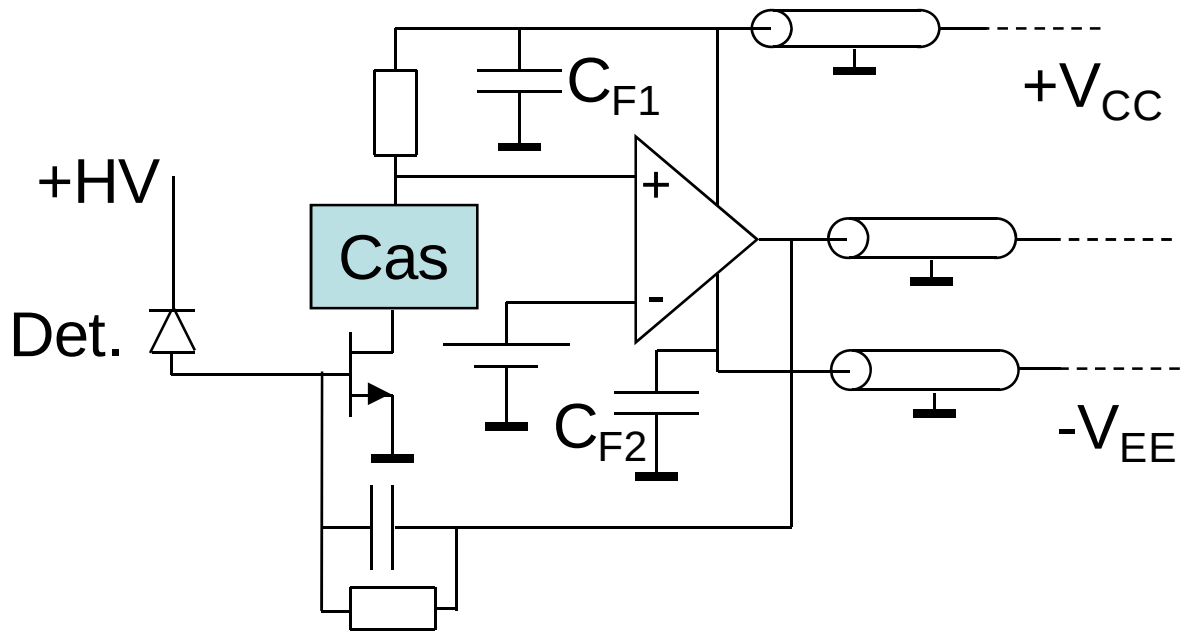

**Figure 1:** The classical Charge Sensitive Preamplifier configuration.

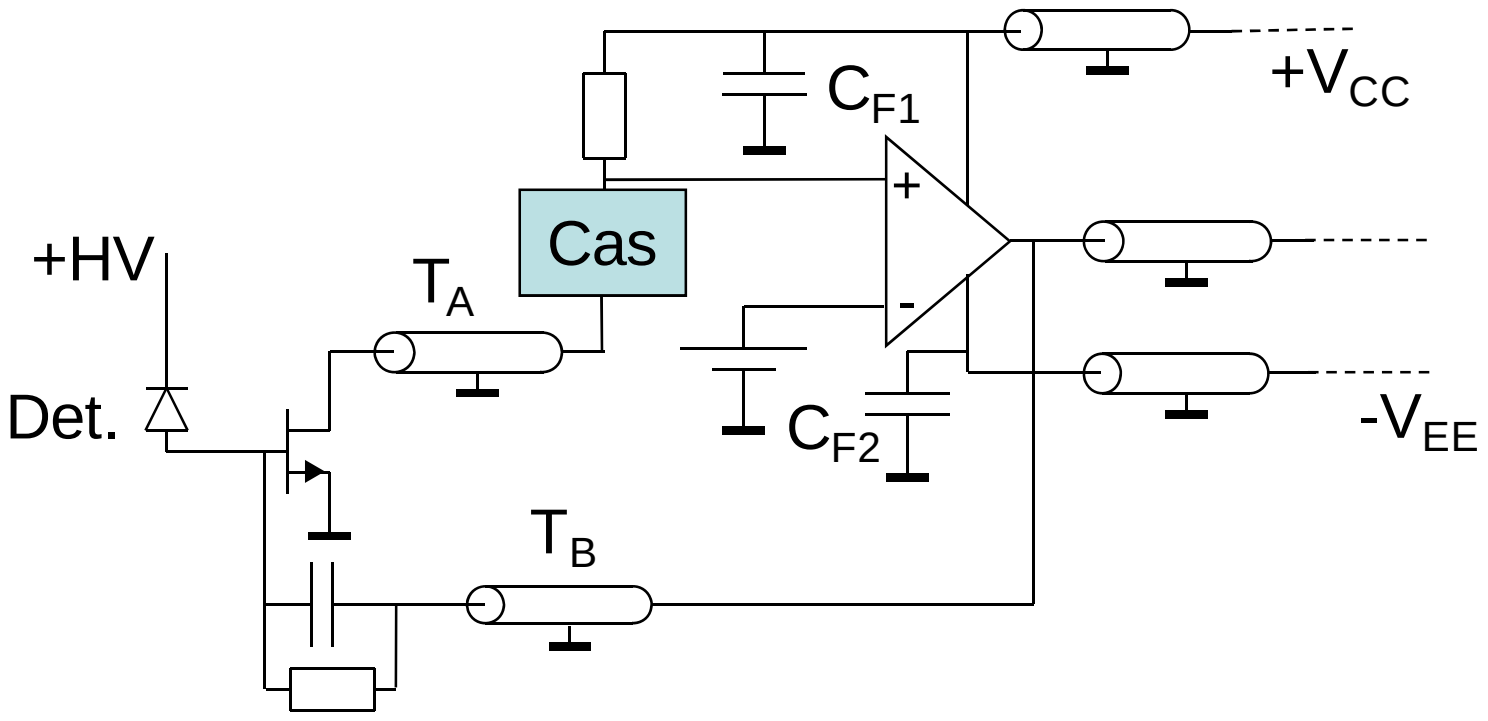

**Figure 2:** The Charge Sensitive Preamplifier of **Figure 1** has now the second stage in a remote location.

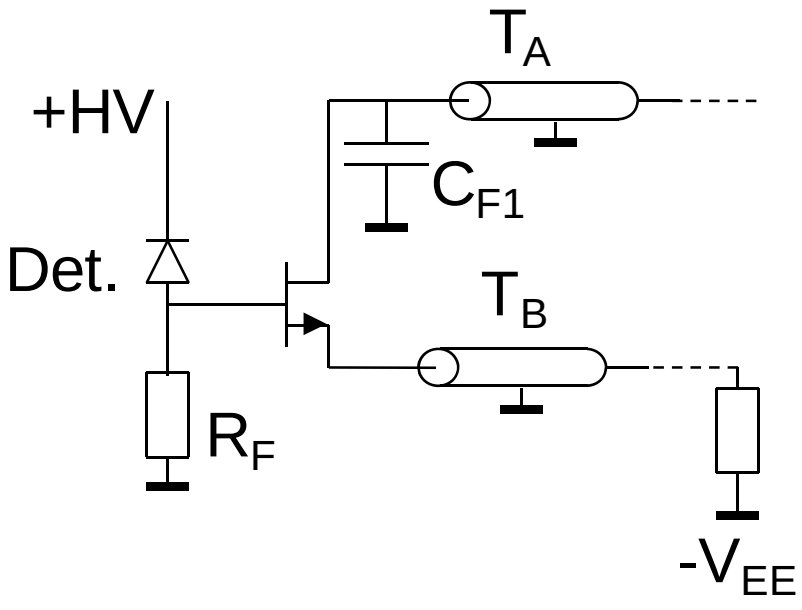

**Figure 3:** The voltage follower readout.

GeFRO, the new readout solution we are proposing, is a merging of the above described set-ups with cross-talk, power dissipation and signal reflection solved. It consist in a fast open loop operated JFET transistor in combination with a slow closed loop network that guarantees the biasing and signal discharge of the input node. In addition it avoids the need of feedback resistor, whose presence is a problem in some low background experiments. The schematic of GeFRO is in Figure 4. The fast open loop path is composed of the common source JFET $J_{IN}$ followed by the room temperature operated feed-backed second stage with gain B. The fast detector pulse is integrated across the capacitance present at the JFET input. The gate signal is converted to the channel current by $J_{IN}$, that drives the terminated (AC coupled with $C_T$, $R_T$) coaxial line $T_A$.

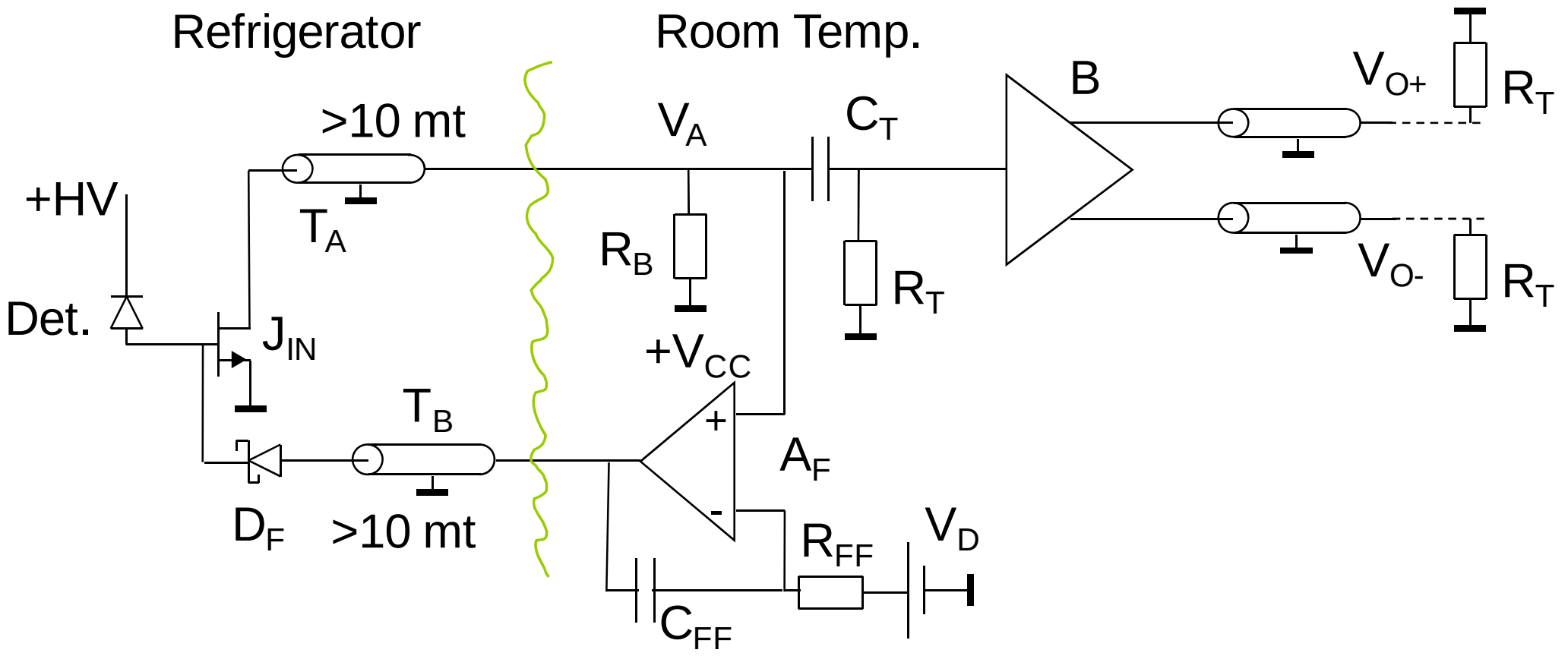

**Figure 4:** GeFRO, the new readout circuit solution.

Neglecting the marginal effect coming from the presence of the drain to gate capacitance, $T_A$ is driven by a current, suppressing any signal loss all along it. In addition, biasing current to $J_{IN}$ is received through $T_A$ itself: signal reflection and signal cross-talk across the supply lines are not present. Power dissipation is minimized since only the JFET dissipates inside the refrigerator. JFET power consumption can be particularly small since it can be operated at small drain to source voltage, $V_{DS}$, because its output impedance, inversely proportional to $V_{DS}$, is in parallel to the small resistor $R_T$ (50-100 Ω, depending on the characteristic of the line).

$J_{IN}$ is operated open loop, at moderate frequencies. The output of the amplifying chain in response to an input charge $Q_D$, neglecting for the moment the DC restoration path $A_F$-$D_F$, is:

$$V_{o+} - V_{o-} = -g_m(V_G)R_T B \frac{Q_D}{C_D + C_{FE} + C_{ch}(V_G)} 1(t) . \quad \textbf{(1)}$$

In (1) $C_D$ is the detector capacitance, $C_{FE} = (1+g_m R_T)C_F$ will be defined in (4), and $C_{ch}(V_G)$ is $J_{IN}$ channel capacitance, $C_{ch}=C_{GS}+C_{GD}$. $C_{ch}$ is the capacitance resulting from the gate to channel junction biased in reverse mode. It is dependent on the width of the space charge region, like the channel current to which it results proportional: Figure 5 is an example of the channel to current dependence of the BF862 n-channel JFET from NXP **[15]**. Transconductance $g_m(V_G)$ depends on both the gate voltage and temperature. Experiments to be operated in a cryogenic environment have stable the temperature and $g_m$ exploits this opportunity. Otherwise it is possible to maintain $g_m$ tuned by allowing the reference voltage $V_D$ to vary with ambient temperature properly. Channel current vs gate voltage and transconductance vs channel current at liquid nitrogen temperature (LN2, 77 K) are shown in Figure 6 and Figure 7. As it can be seen, gain may depart from linearity and shows a tiny quadratic behavior depending on the expected energy dynamic range of the detector. As an example a dynamic range of 10 MeV in a standard Ge coaxial detector of 30 pF capacitance, results in about 10 mV of signal (in matching conditions) at the $J_{IN}$ input, a very small excursion. The same energy range in a Broad Energy

Germanium (BEGe) **[10]** or BEGe-like (pin-point contact) detector, characterized by about 1 pF of detector capacitance, results in about 0.25 V at the JFET input gate in matching conditions and the response will show a quadratic effect. Figure 8 is the response obtained with a BF862 JFET from NXP, connected to a simulated detector with a negligible capacitance (BEGe like detector). The range explored was up to 1 $Me^{-}$, corresponding to about 3 MeV released by an impinging particle in the Ge detector. The maximum integral linear error found was 0.25%.

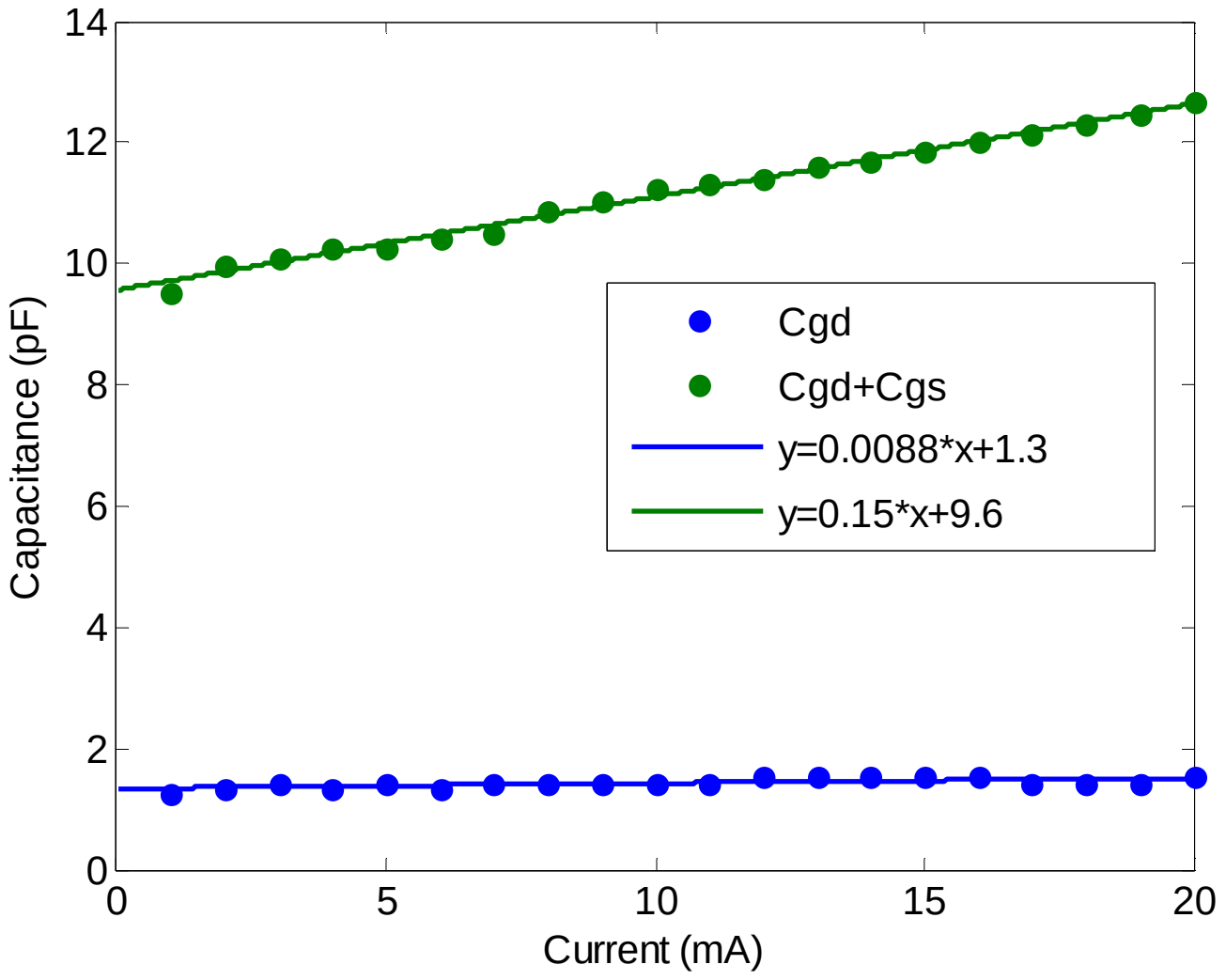

**Figure 5:** Channel Capacitance, $C_{GS}$+$C_{GD}$, and drain to gate capacitance, $C_{GD}$ vs the channel current for a BF862. Drain to source voltage was fixed to 2 V.

The DC stabilization network of GeFRO in Figure 4, $A_F$-$D_F$, is a slow closed loop path that is unaffected from the phase shift that the long cabling connections can add. It sets the drain voltage to $V_D$ and the channel current of $J_{IN}$ by adjusting its gate potential properly. The DC restoration of the feedback path is done exploiting a Schottky diode, $D_F$, in reverse biasing condition. This device was found to be excellent for this purpose because it behaves at cryogenic temperature like a non linear resistor in parallel to a capacitance of less than 1 pF. This way a large range of detector currents can be sustained. As an example, a detector current of the order of 10 pA results in a working point for $D_F$ corresponding to a dynamic resistance of a few GΩ, Figure 9. Alternatively, a standard solution with a high value resistor can be considered instead.

The readout of the DC voltage drop across the Schottky diode (or the feedback resistor) is a valuable feature when working with Ge detectors, as it allows the measurement of the actual detector leakage current.

In a feed-backed network the signal is proportional to the loop gain T (=gain×feedback return path β), a decreasing function, according to -T/β(1-T), where 1/β is the gain in the ideal condition of T=∞. Stability becomes critical as long as T approaches 1, or its phase shifts by π at high frequencies. Divergence is avoided if |T| is less than one at the π phase shift. A very good stability is obtained if |T| is less than 1 at adequate low frequencies. This was the philosophy we adopted in the design of our readout. If we cut the loop of Figure 4 at the input of $A_F$ (so avoiding to perturb the network) and apply a test signal to be propagated back to the output we obtain, once that the impedance of $D_F$ is approximated with a capacitance $C_F$:

$$T = -\frac{g_m R_T C_F}{C_D + C_{ch} + C_F} \frac{1 + sC_{FF}R_{FF}}{sC_{FF}R_{FF}} . \qquad \textbf{(2)}$$

In (2) choosing the time constant $C_{FF}R_{FF}$ large enough, a value of |T| smaller than 1 easily results when $C_F$ is smaller than $C_D$+$C_{ch}$ and $g_mR_T$ is a few units. If the connecting coaxial cables are 10 m long a delay of about 100 ns is expected (signal back and forward) so that $C_{FF}R_{FF}$ should be not smaller than about 1 μs. The output signal at node $V_A$ of Figure 4 is now given by:

$$\begin{aligned} V_A &= \frac{1}{\beta}\frac{-T}{1-T} = -\frac{C_{FF}R_{FF}}{1+sC_{FF}R_{FF}}\frac{-T}{1-T}\frac{Q_D}{C_F} \\ &= -\frac{g_m R_T C_F}{C_D + C_{ch} + (1+g_m R_T)C_F}\frac{1}{s+\frac{1}{C_{FF}R_{FF}}\frac{g_m R_T C_F}{C_D+C_{ch}+(1+g_m R_T)C_F}}\frac{Q_D}{C_F} \qquad \textbf{(3)} \\ &= -\frac{T_o}{s+\frac{T_o}{C_{FF}R_{FF}}}\frac{Q_D}{C_F}, \qquad \left(T_o = \frac{g_m R_T C_F}{C_D + C_{ch} + (1+g_m R_T)C_F}\right). \end{aligned}$$

Parameter $T_o$ in the above equation is less than 1 and takes into account the wanted departing from the ideal behaviour of the loop gain. Coming back to time domain at the readout output from (3):

$$\begin{aligned} V_{o+} - V_{o-} &= -BT_o \exp\left(-\frac{t}{C_{FF}R_{FF}/T_o}\right)1(t)\frac{Q_D}{C_F} \\ &\approx -Bg_m R_T \exp\left(-\frac{t}{C_{FF}R_{FF}(C_D+C_{ch}+(1+g_m R_T)C_F)/g_m R_T C_F}\right)\frac{1(t)Q_D}{C_D + C_{ch} + +(1+g_m R_T)C_F} . \end{aligned} \qquad \textbf{(4)}$$

The product $g_mR_T$ in (4) is close to one, or a few times one. As a consequence the signal is mainly inversely proportional to the input capacitance $C_D$+$C_{ch}$, while capacitance $C_{FE}$ of (1) results now in (1+$g_mR_T$)$C_F$. The fall time of the signal can be shaped by suitable setting of the time constant $C_{FF}R_{FF}$. This is an useful property since it can be small enough to minimize the microphonic effects coming from cryogenic liquid boil-off, and large enough to keep the noise from $A_F$ at a small level, as it will be discussed below.

In summary, the behavior of the readout can be described by the amplification of a charge signal developed across an (open loop) input capacitance, $C_D$+$C_{ch}$, and with a (open loop) gain $g_mR_T$; DC restoration of the baseline is made by a slow feedback path.

The upper limit in frequency response is not considered in (4), and in the actual circuit the bandwidth is settled by the second stage B of Figure 4. In Figure 10 one can see that the speed of response can be very fast, a few ns, after an accurate choice of the second stage B is made.

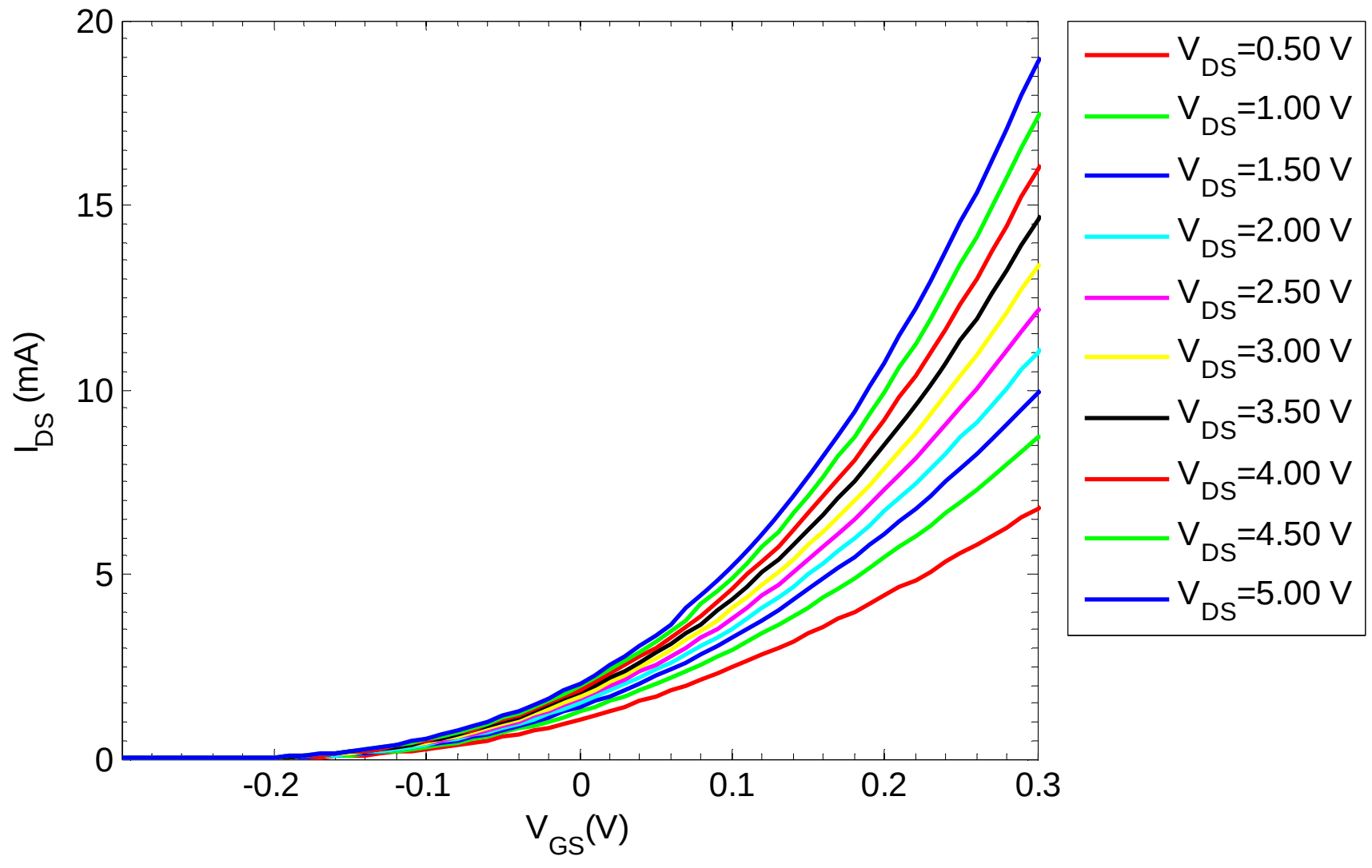

**Figure 6:** Channel current vs gate voltage for the BF862 at LN2 temperature.

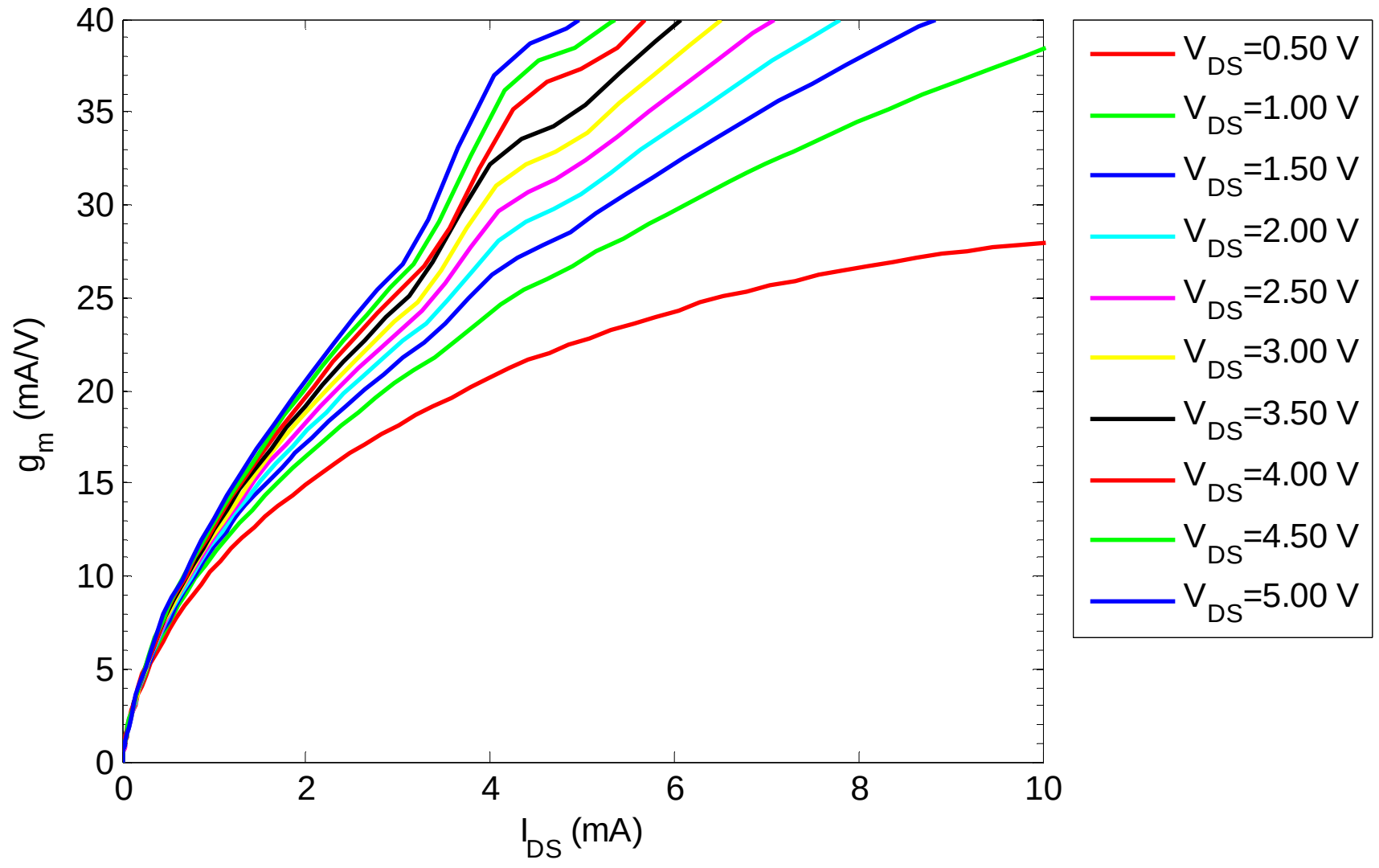

**Figure 7:** Transconductance vs channel current for the BF862 at LN2 temperature.

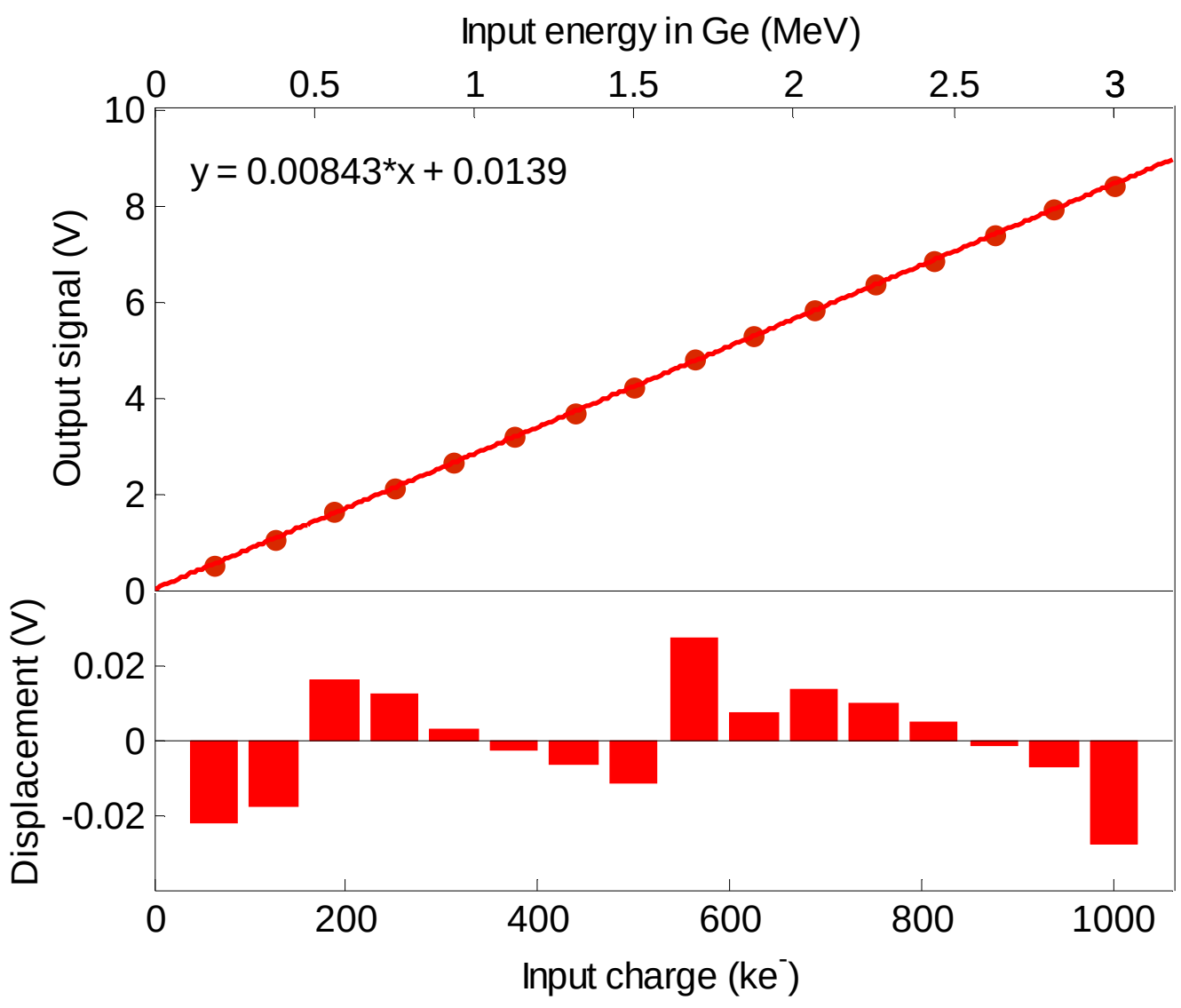

**Figure 8:** Response to 0 - 1 Me$^-$ (~ 0 - 3 MeV released in germanium) charge range with a simulated detector of negligible capacitance; as JFET we used the BF862 from NXP.

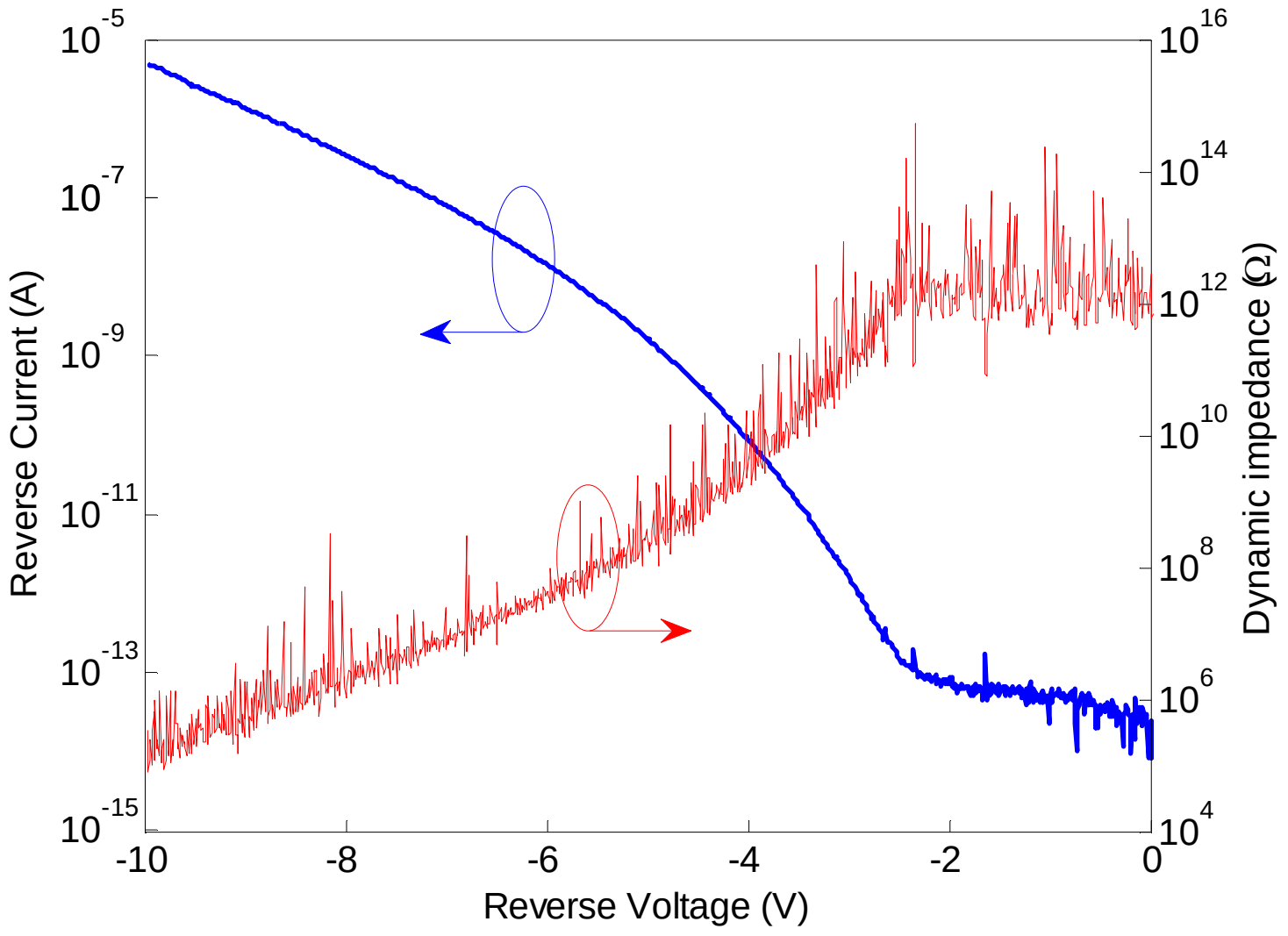

**Figure 9:** Reverse current (absolute value, left axis) and dynamic impedance (right axis) vs the reverse voltage for the Schottky diode BAT17 at LN2 temperature. The calibration of the V-I characteristic of the diode allows the readout of the actual detector leakage current.

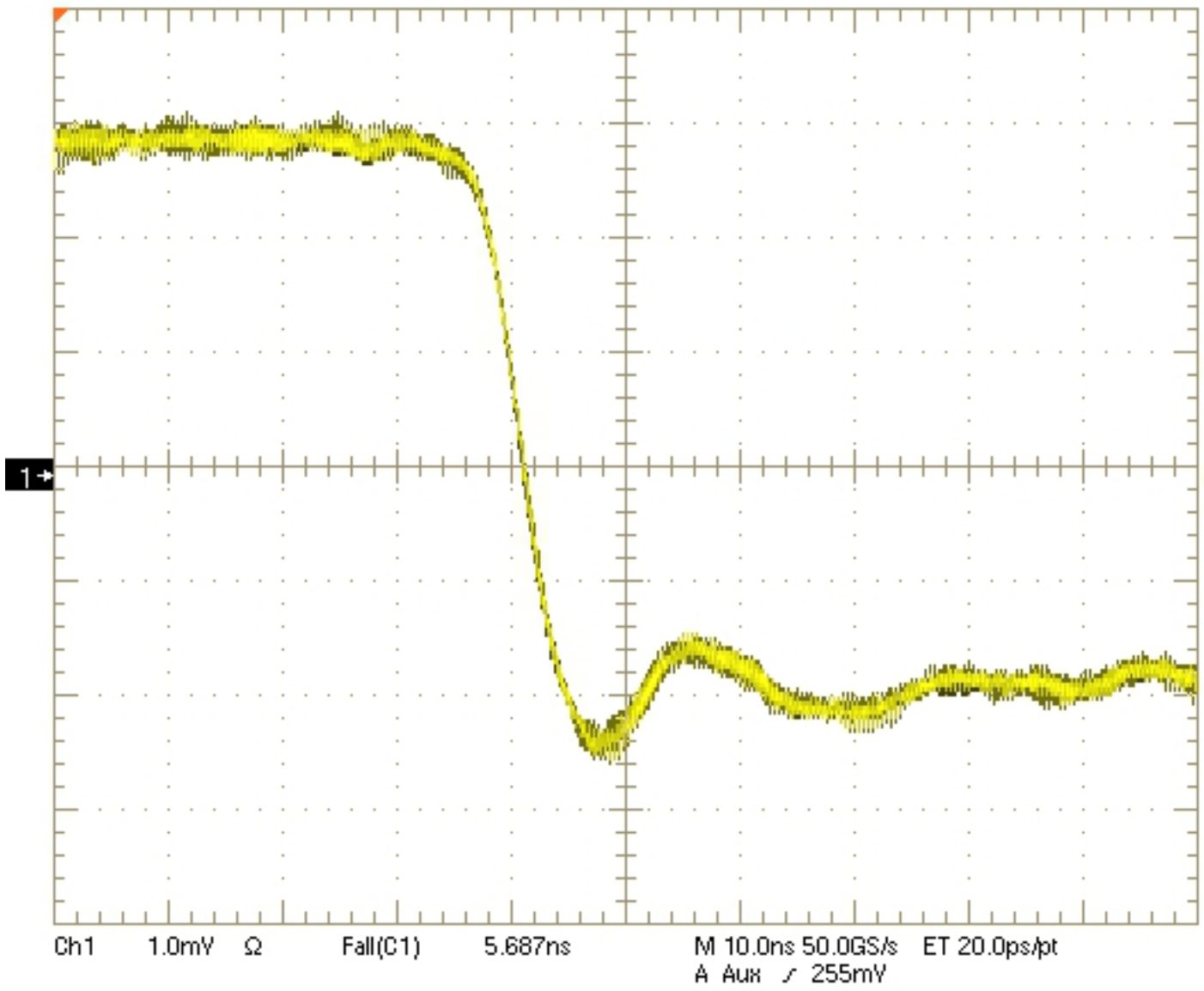

**Figure 10:** The signal response of GeFRO to 100 $Ke^-$ input charge; as second stage B, a pair of AD811 was adopted. Horizontal scale is 10 ns/div, vertical scale 1 mV/div. The fall time of the signal is 5.7 ns.

## 3. Noise performances

The gain of the proposed readout is proportional to the parameter $T_o$ from (3) which is smaller than 1, in any conditions. As a consequence, care must be put in the selection of the second stage B and amplifier $A_F$. The termination resistor $R_T$ could also contribute to the noise performance. If we suppose to filter the output of GeFRO with a standard CR-$RC^n$ or Gaussian-like filter we expect to find for the Equivalent Noise Charge, ENC, or the RMS charge to be injected at the input to make unity the signal to noise ratio (see Appendix for the detailed calculations):

$$
\begin{aligned}
ENC^2 \approx & \left(C_D + C_{ch} + C_F\right)^2 \cdot \\
& \cdot \left\{ \frac{\alpha}{\tau} \left[ \overline{e_{INw}^2} + \left( \frac{C_F}{C_D + C_{ch} + C_F} \right)^2 \left( 1 + \left( \frac{\tau}{C_{FF} R_{FF}} \right)^2 \frac{\gamma}{\alpha} \right) \overline{e_{AF}^2} \right] + \beta A_{INf} \right\} \\
& + \frac{\alpha \left(C_D + C_{ch} + C_F\right)^2}{\tau} \cdot \\
& \cdot \left[ \left( \frac{C_D + C_{ch} + \left(1 + g_m R_T\right) C_F}{C_D + C_{ch} + C_F} \right)^2 \frac{\overline{e_{II}^2}}{\left(g_m R_T\right)^2} + \frac{R_T^2 \left( \overline{i_{II}^2} + \overline{i_{AF}^2} \right) + 4 K_B T R_T}{\left(g_m R_T\right)^2} \right] \\
& + \gamma \tau 2 q 2 I_D .
\end{aligned} \qquad (5)
$$

In (5) α, β and γ are the coefficients for the series white, series 1/f and parallel noise that depend on the adopted filter, while τ is its shaping time constant (as an instance for a CR-$RC^{10}$ filter it results 0.15, 3.23 and 2.97 for α, β and γ, respectively); with reference to Figure 11 $\overline{e_{INw}^2}$ and $A_{INf}$ are the white and 1/f noise sources of $J_{IN}$, strongly dependent on temperature at cold **[16]**,

**[17]**; $\overline{e_{II}^2}$ and $\overline{i_{II}^2}$ are the series and parallel noise of the output amplifier B and $\overline{e_{A_F}^2}$ and $\overline{i_{A_F}^2}$ those of amplifier $A_F$, respectively.

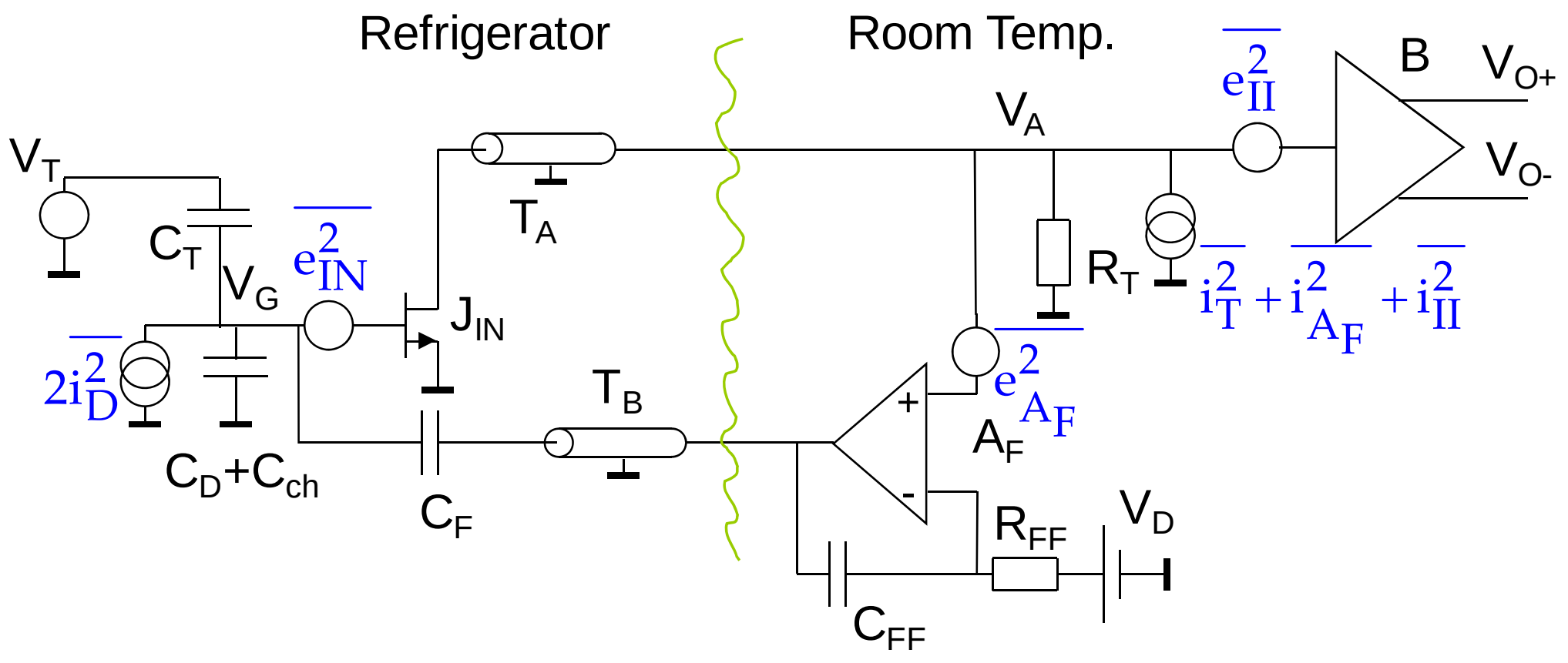

**Figure 11:** The main noise sources of GeFRO for noise calculation.

We have characterized some different JFETs in the GeFRO configuration of Figure 4 at LN2 temperature. Measurements have been taken using as a second stage B a very low noise amplifier, whose noise figure is < 0.5 nV/√Hz, designed for a bolometric application **[18]**. An Ortec 762 semi-Gaussian filter and a 30 MHz bandwidth true RMS voltmeter URE3 from Rohde & Schwarz complete the setup.

As an application example, we show the noise results obtained with the BF862 and the BAT17 as input JFET and Schottky diode of Figure 4 respectively. The circuit was submerged in LN2. In this case the parallel noise was completely negligible and the ENC was expected to decrease with the shaping time constant of the filter except for a constant contribution given by the 1/f noise. In Figure 12 the results from a set of measurements is shown. Three different channel currents for the input JFET have been selected, while the input capacitance was set to a negligible value, BEGe case, and 33 pF, typical for a standard Ge detector. As it can be seen the noise is slightly dependent on channel current at large shaping time since 1/f noise takes over. The energy resolution is given also in $eV_{FWHM}$ (Full Width Half Maximum) converted to a Ge detector on the right axis.

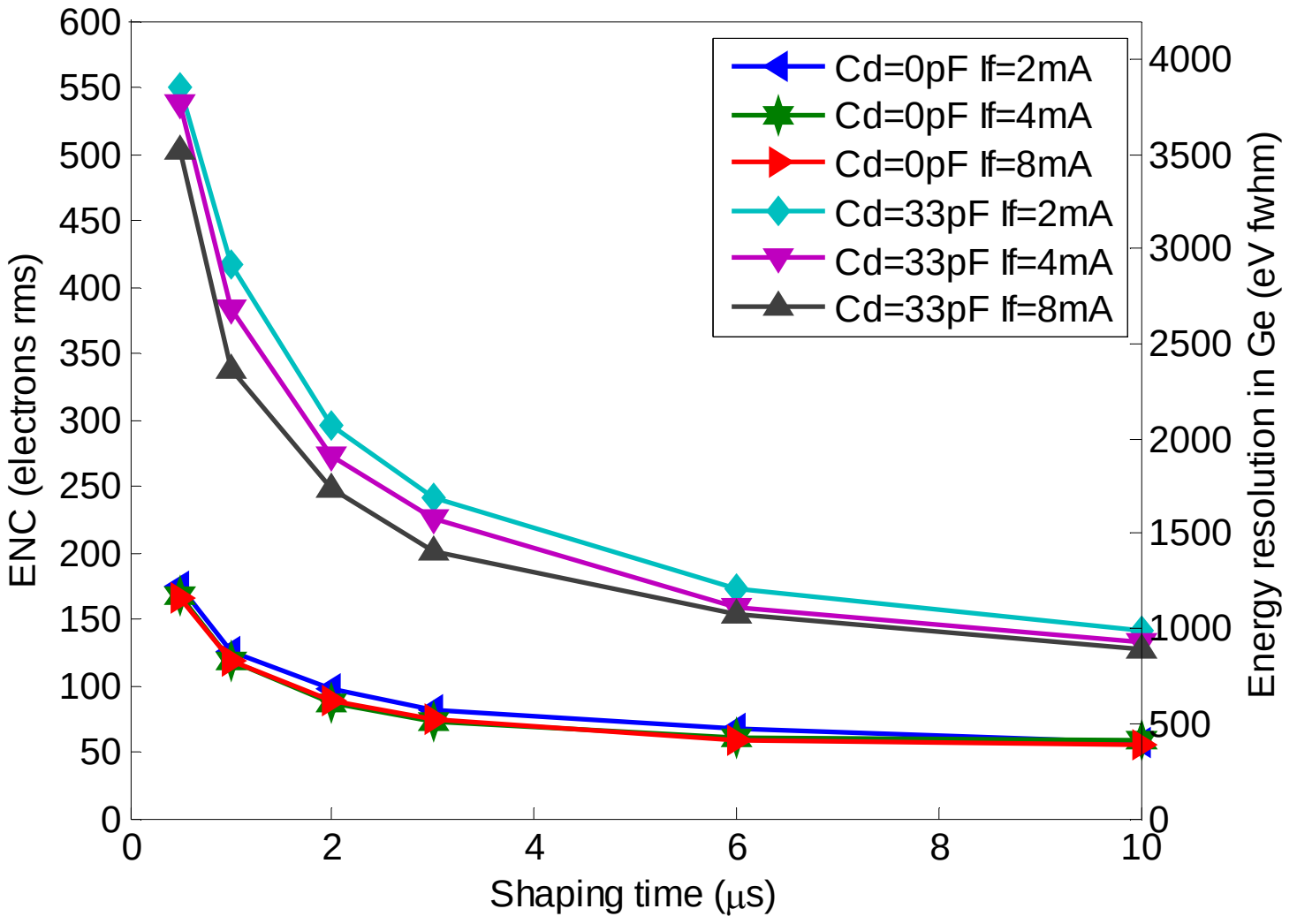

**Figure 12:** ENC vs the shaping time (left axis) of GeFRO**Figure 4**; the input JFET is a BF862 and the Schottky diode a BAT17. Measurements have been taken at LN2 temperature. On the right axis the corresponding energy resolution for a Ge detector in FWHM is given. Measurements have been taken at different working point of the BF862 JFET ($I_{DS}$ ranging from 2 mA to 8 mA) and simulated detector capacitances.

The proposed solution has been tested with a Ge detector illuminated with a $^{22}$Na $\gamma$ source ($E_\gamma$ = 1.275 MeV). Figure 13 shows the results obtained at 10 μs shaping time. The resolution at 1.2 MeV is 2.44 $KeV_{FWHM}$, limited mainly by the detector and interference noise present in the laboratory environment. A proof of this interpretation is the spectrum of Figure 14, obtained in the same conditions but the shaping time that was set to 6 μs this time. The resolution improved to 2.21 $KeV_{FWHM}$ although the larger preamplifier expected noise. The measured detector leakage current was 10 pA, not expected to contribute to the performances in both conditions.

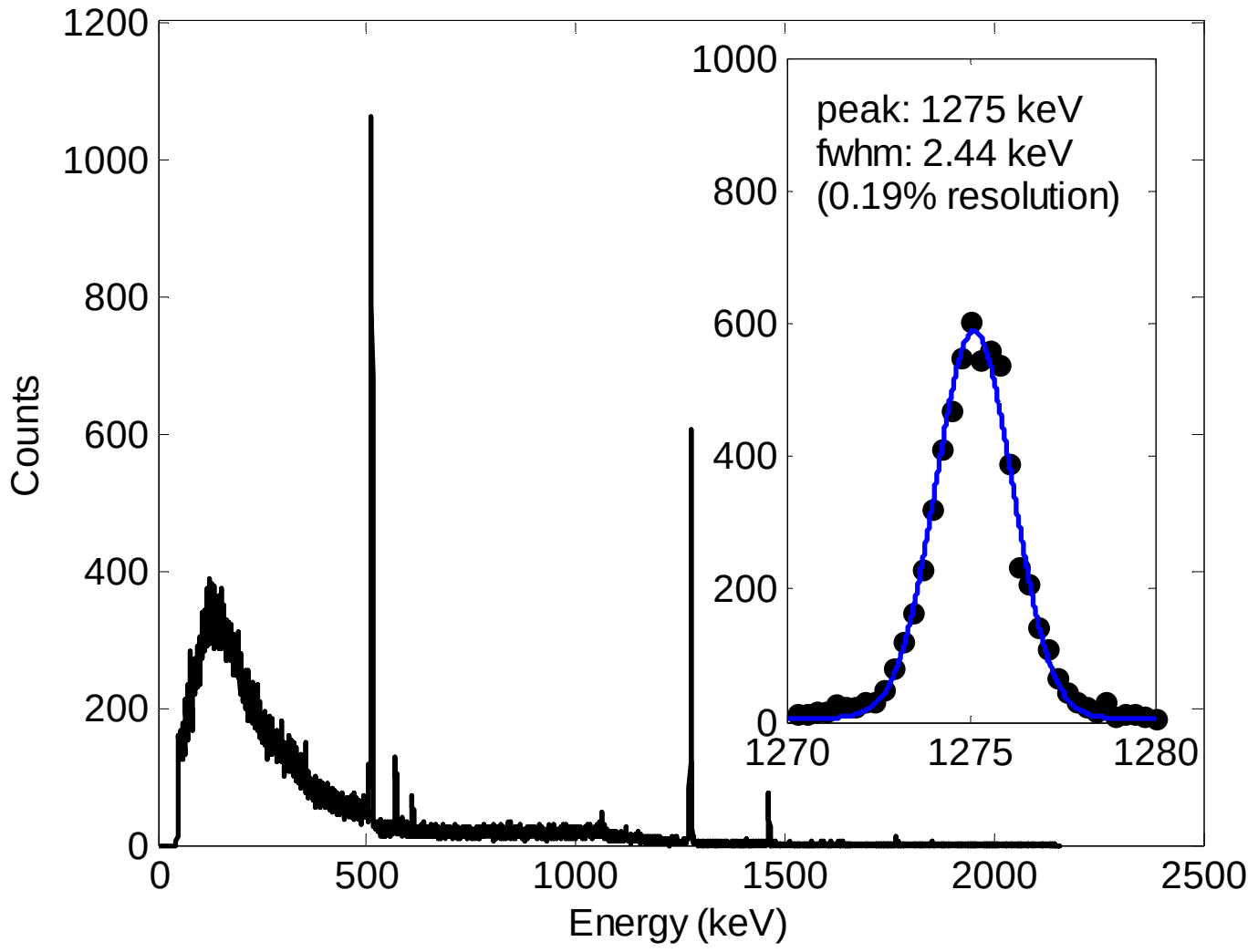

**Figure 13:** The measured energy spectrum, at 10 μs shaping time, when a Ge coaxial detector ($C_{det}$~ 20 pF) is readout by the GeFRO circuit and irradiated by a $^{22}$Na $\gamma$ source. The achieved resolution is 2.44 keV FWHM at the 1.275 MeV $\gamma$ line, corresponding to 0.19 %.

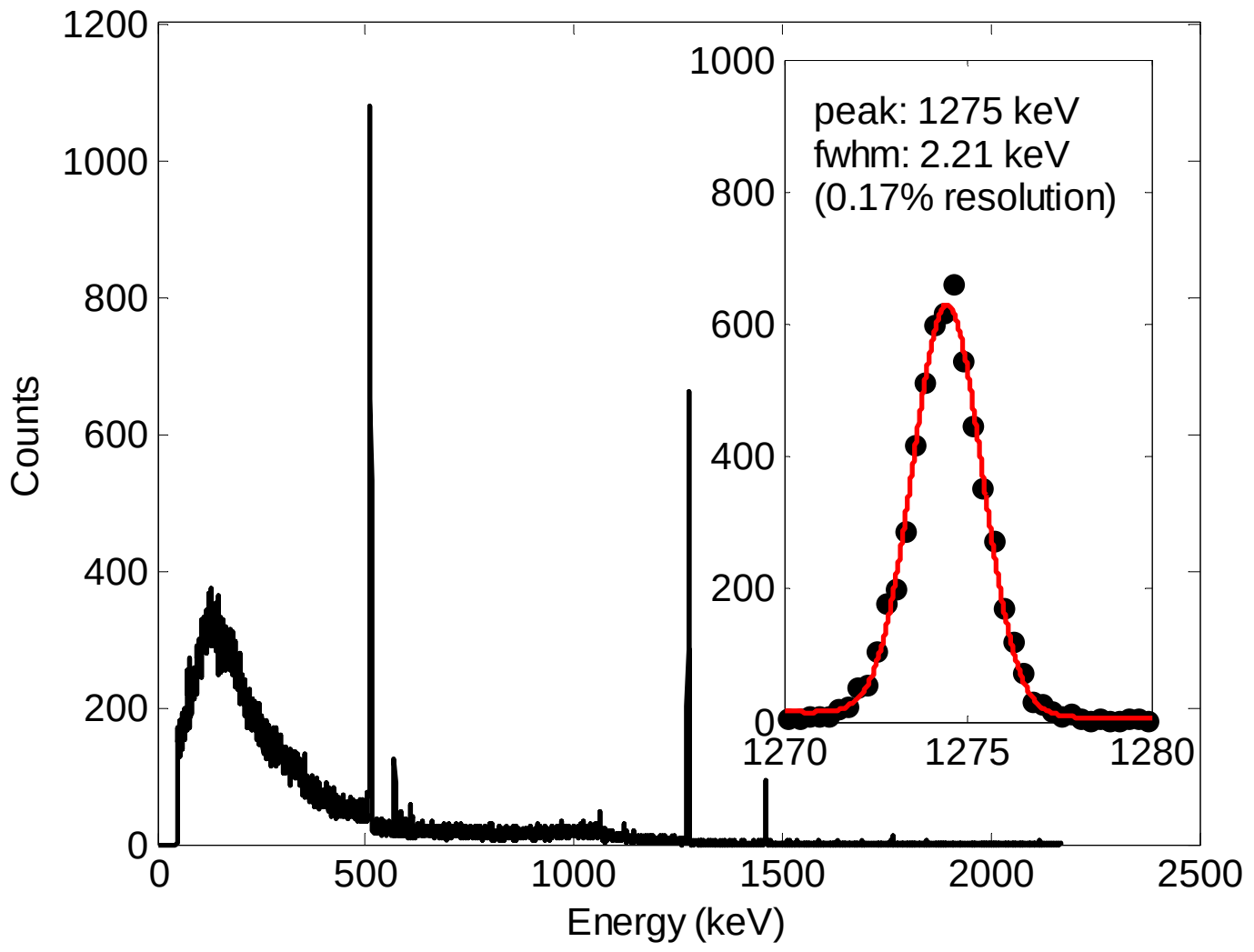

**Figure 14:** The measured energy spectrum, at 6 μs shaping time, when a Ge coaxial detector ($C_{det}$~ 20 pF) is readout by the GeFRO circuit and irradiated by a $^{22}Na$ γ source. The achieved resolution is 2.21 keV FWHM at the 1.275 MeV γ line, corresponding to 0.17%.

## 4. Conclusions

GeFRO, a new approach to the readout of ionization detectors operated at cold, is given. It uses only one JFET transistor and a diode for DC restoration and operates in close proximity of the detector, allowing to minimize space occupation; this is a valuable feature to read-out detectors operated in low-background experiments. The circuit is able to drive a terminated coaxial cable and guarantees high speed and low noise. Power dissipation is minimal thanks to the fact that only the JFET dissipates ad cold. The circuit is being tested at the GERDA collaboration.

## Acknowledgments

We thank the invaluable technical support from Antonio De Lucia, from the Università di Milano Bicocca, in taking several of the measurements shown in this paper.

## 5. Appendix

To measure the noise the calibration of the network must be made first, by simulating the detector input charge. If a detector event, a very short charge signal, is input to GeFRO in Figure 11 a voltage is generated equal to:

$$V_G \approx \frac{Q_D}{sC_{Tot}}. \tag{6}$$

Where $C_{Tot}$ is the overall capacitance seen at $J_{IN}$ input node. If now a step voltage $V_T$ is sent to the test capacitance $C_T$ we obtain:

$$V_G \approx \frac{C_T V_T}{sC_{Tot}}. \tag{7}$$

Therefore the test calibrating charge results $C_T V_T$.

In Figure 11 the main sources of noise are shown. The evaluation of the contribution of all the sources that are in the feedback loop is easy if we suppose first T>>1, evaluate the corresponding transfer function and then multiply it by –T/(1-T). For the case the time constant of the exponential term of (4) is much larger than the shaping time constant τ we can do the approximation:

$$\frac{-T}{1-T}=\frac{T_o\left(1+sC_{FF}R_{FF}\right)}{C_{FF}R_{FF}\left(s+\frac{T_o}{C_{FF}R_{FF}}\right)}\approx\frac{T_o\left(1+sC_{FF}R_{FF}\right)}{sC_{FF}R_{FF}}. \tag{8}$$

The response of the system to the calibrating charge is:

$$V_o=-B\frac{C_{FF}R_{FF}}{1+sC_{FF}R_{FF}}\frac{-T}{1-T}\frac{C_TV_T}{C_F}\approx -B\frac{T_o}{sC_F}C_TV_T. \tag{9}$$

When the signal (9) is sent to the input of the shaper, the resulting output has a peaking time where the maximum is found. This maximum does not depend on the shaping time constant, but only in the shape of the filter. Let's call it $f_{MAX}$. The ratio between the measured $V_{oMAX}$ in response to the test input charge $C_TV_T$ allows to determine the transfer function for the signal.

$$TF=\frac{V_{oMAX}}{C_TV_T}=-\frac{BT_o}{C_F}f_{MAX}. \tag{10}$$

Let us now evaluate the noise. Noise from $\overline{e_{IN}^2}$ and $\overline{e_{A_F}^2}$ results in:

$$\begin{aligned}\overline{V_{o1}^2}&=B^2\left[\left|\frac{sC_{FF}R_{FF}}{1+sC_{FF}R_{FF}}\frac{C_D+C_{ch}+C_F}{C_F}\right|^2\overline{e_{IN}^2}+\overline{e_{A_F}^2}\right]\left|\frac{-T}{1-T}\right|^2\\&=\left[\left|\frac{sC_{FF}R_{FF}}{1+sC_{FF}R_{FF}}\frac{C_D+C_{ch}+C_F}{C_F}\right|^2\left(\overline{e_{INw}^2}+\frac{A_{INf}}{f}\right)+\overline{e_{A_F}^2}\right]\left|\frac{-T}{1-T}\right|^2.\end{aligned} \tag{11}$$

Noise from input parallel noise is:

$$\overline{V_{o2}^2}=B^2\left|\frac{1}{1+sC_{FF}R_{FF}}\frac{C_{FF}R_{FF}}{C_F}\frac{-T}{1-T}\right|^2 2q2I_D. \tag{12}$$

Parallel noise of $A_F$, $R_T$ and amplifier B is developed across the output impedance present at node $V_A$, which is $R_T/(1-T)$:

$$\overline{V_{o3}^2}=B^2\left|\frac{R_T}{1-T}\right|^2\left(\overline{i_T^2}+\overline{i_{A_F}^2}+\overline{i_{II}^2}\right)\approx\frac{B^2T_o^2R_T^2}{\left(g_mR_T\right)^2}\left(C_D+C_{ch}+C_F\right)^2\left(\overline{i_T^2}+\overline{i_{A_F}^2}+\overline{i_{II}^2}\right). \tag{13}$$

Therefore, considering also the input series noise of amplifier B:

$$\begin{aligned}\overline{V_o^2}&=\overline{V_{o1}^2}+\overline{V_{o2}^2}+\overline{V_{o3}^2}+B^2\overline{e_{II}^2}\\&=B^2T_o^2\left[\left(\frac{C_D+C_{ch}+C_F}{C_F}\right)^2\left(\overline{e_{INw}^2}+\frac{A_{INf}}{f}\right)+\frac{1+\left(\omega C_{FF}R_{FF}\right)^2}{\left(\omega C_{FF}R_{FF}\right)^2}\overline{e_{A_F}^2}\right]\\&+\frac{B^2T_o^2}{\omega^2C_F^2}2q2I_D+\frac{B^2T_o^2R_T^2}{\left(g_mR_T\right)^2}\left(\frac{C_D+C_{ch}+C_F}{C_F}\right)^2\left(\overline{i_T^2}+\overline{i_{A_F}^2}+\overline{i_{II}^2}\right)+B^2\overline{e_{II}^2}.\end{aligned} \tag{14}$$

The maximum of the signal from (10) is independent of frequency. We can therefore define the ENC at any frequency as:

$$\mathrm{ENC}(\omega)=\frac{\overline{V_o^2}}{TF^2}\left|\frac{s\tau}{(1+s\tau)^n}\right|^2=C_F^2\frac{\overline{V_o^2}}{(BT_o f_{MAX})^2}\left|\frac{s\tau}{(1+s\tau)^n}\right|^2$$

$$=\left\{(C_D+C_{ch}+C_F)^2\left(\overline{e_{INw}^2}+\frac{A_{INf}}{f}\right)+C_F^2\frac{1+(\omega C_{FF}R_{FF})^2}{(\omega C_{FF}R_{FF})^2}\overline{e_{AF}^2}\right\}\left|\frac{s\tau}{(1+s\tau)^n f_{MAX}}\right|^2 \qquad \textbf{(15)}$$

$$+\left\{\frac{1}{\omega^2}2q2I_D+\frac{(C_D+C_{ch}+C_F)^2R_T^2}{(g_mR_T)^2}\left(\overline{i_T^2}+\overline{i_{AF}^2}+\overline{i_{II}^2}\right)+C_F^2\frac{\overline{e_{II}^2}}{T_o^2}\right\}\left|\frac{s\tau}{(1+s\tau)^n f_{MAX}}\right|^2 .$$

In the above equation we have considered a CR-RC$^n$ shaper. Integrating ENC(ω) results in (5) once an obvious meaning of the terms is considered.